\documentclass[12pt]{article}

\usepackage{scicite}

\usepackage{times}

\usepackage{color}
\usepackage{graphicx}

\newenvironment{sciabstract}{%
\begin{quote} \bf}
{\end{quote}}

\newcounter{lastnote}
\newenvironment{scilastnote}{%
\setcounter{lastnote}{\value{enumiv}}%
\addtocounter{lastnote}{+1}%
\begin{list}%
{\arabic{lastnote}.}
{\setlength{\leftmargin}{.22in}}
{\setlength{\labelsep}{.5em}}}
{\end{list}}

\title{Observation of many-body long-range tunneling after a quantum quench}

\author
{Florian Meinert,$^{1}$ Manfred J. Mark,$^{1}$ Emil Kirilov,$^{1}$ Katharina Lauber,$^{1}$ \\
Philipp Weinmann,$^{1}$ Michael Gr\"obner,$^{1}$ Andrew J. Daley,$^{2,3}$ \\
Hanns-Christoph N\"agerl$^{1\ast}$ \\
\\
\normalsize{$^{1}$Institut f\"ur Experimentalphysik und Zentrum f\"ur Quantenphysik,} \\
\normalsize{Universit\"at Innsbruck, 6020 Innsbruck, Austria}\\
\normalsize{$^{2}$Department of Physics and Astronomy, University of Pittsburgh, Pittsburgh, PA 15260, USA}\\
\normalsize{$^{3}$Department of Physics and SUPA, University of Strathclyde, Glasgow G4 0NG, UK}\\
\\
\normalsize{$^\ast$To whom correspondence should be addressed; E-mail:  Christoph.Naegerl@uibk.ac.at}
}

\date{}

\begin{document}


\baselineskip24pt


\maketitle


\begin{sciabstract}
  Quantum tunneling constitutes one of the most fundamental processes in nature. We observe resonantly-enhanced long-range quantum tunneling in one-dimen\-sional Mott-insulating Hubbard chains that are suddenly quenched into a tilted configuration. Higher-order many-body tunneling processes occur over up to five lattice sites when the tilt per site is tuned to integer fractions of the Mott gap. Starting from a one-atom-per-site Mott state the response of the many-body quantum system is observed as resonances in the number of doubly occupied sites and in the emerging coherence in momentum space. Second- and third-order tunneling shows up in the transient response after the tilt, from which we extract the characteristic scaling in accordance with perturbation theory and numerical simulations.
\end{sciabstract}


Quantum mechanics allows particles to overcome barriers even when the particles' energies are not sufficient for a classical jump. Quantum tunneling is ubiquitous in physics and forms the basis for a multitude of fundamental effects related to electronic transport, nuclear motion, and light propagation \cite{Ankerhold2010}. In lattice systems, for example in the context of solid state and condensed matter systems, tunneling as characterized by a rate $J$ usually constitutes a sequential and individual process, where the particles move from one site to the next in an uncorrelated way. Such motion is at the heart of superfluidity in weakly-interacting lattice systems \cite{Morsch2006}. Strong interactions, however, are capable of changing the situation completely. The particles' dynamics become correlated, i.e. the motion of a particle is conditioned upon other particles' motion, thereby giving rise to effective long-range interactions beyond the local on-site interactions. Correlated tunneling processes are believed to play an important role, for example, in superconductivity of the celebrated cuprate systems \cite{MacDonald1988,Anderson2004,Norman2011}. Second-order tunneling involving two particles in double wells has been directly observed in cold atom experiments \cite{Foelling2007}. That process results in an effective nearest-neighbor super-exchange interaction \cite{Trotzky2008,Greif2013}, which forms the basis of important forms of quantum magnetism \cite{Dagotto1994}, and provides a starting point for the formation of
complex quantum many-body phases, as well as dynamical buildup of massive quantum entanglement in a many-body system \cite{Amico2008,Trotzky2012,Collath2007}. Third- and higher-order tunneling processes are capable of coupling particles across three and more sites, allowing for the generation of yet stronger correlations.

Here, we directly observe dynamics in a quantum many-body system dominated by second- and third-order tunneling and identify resonant fourth- and fifth-order tunneling processes. Previous experiments with cold atoms have studied single-particle tunneling loss via higher band resonances \cite{Sias2007}. We study long-range many-body tunneling processes that we resonantly enhance in tilted one-dimensional (1D) Mott-insulating ``Ising'' chains of bosons \cite{Simon2011,Bakr2011,Sachdev2002,Meinert2013}. Specifically, we perform a quantum quench to a highly non-equilibrium situation by rapidly tilting a one-atom Mott insulator to an integer fraction $U/n$ of the Mott gap $U$. We record the probability for atom-pair formation as a function of time after the quench and find that the rate for pair formation is set by $\alpha_n \times J^n/(U^{n-1}/n)$ for $n=2$ and $3$ with a surprisingly large prefactor $\alpha_n \approx 36$ in accordance with the expected scaling $\propto J^n/U^{n-1}$ from perturbation theory.

Our experiment is based on an array of 1D chains of atoms in an optical lattice near zero temperature \cite{Meinert2013}. We model the system by a single-band Bose-Hubbard (BH) Hamiltonian \cite{Jaksch1998,Methods}. For $U \gg J$ the many-body ground state is a Mott insulator with unit occupation at commensurate filling (Fig.1 A). This phase is characterized by exponentially localized atoms and highly suppressed tunneling. In addition, we superimpose a linear gradient potential, which introduces a site-to-site constant energy shift $E$. Tilting the initial Mott state quickly to $E \approx U/n$ initiates resonant tunneling to the $n-$th neighbor for all sites simultaneously (lower part of Fig.1 A). For $n=1$ one couples to nearest-neighbor dipole states and observes strong coherent oscillations in the number of doubly occupied sites (doublons) with a characteristic frequency $4J$ \cite{Meinert2013}. For $n>1$ resonant tunnel coupling occurs across $n-1$ intermediate lattice sites. The process involves up to $n$ other particles, giving rise to occupation-dependent $n$-th-order tunneling. Since all particles participate in a tunnel process across $n$ sites, one expects the build-up of massive correlations in the interacting many-body system.

We prepare an ensemble of 1D Mott insulators \cite{Meinert2013} starting from a 3D Bose-Einstein condensate (BEC) of typically $8.5 \times 10^4$ Cs atoms without detectable uncondensed fraction. The BEC is levitated against gravity by a magnetic field gradient of $|\nabla B| \approx 31.1 \, \rm{G/cm}$ and initially held in a crossed optical dipole trap \cite{Weber2002,Kraemer2004}. We load the sample adiabatically into a cubic 3D optical lattice generated by laser beams at a wavelength of $\lambda_{\rm{l}} = 1064.5 \, \rm{nm}$, thereby creating a singly-occupied 3D Mott insulator for a lattice depth of $V_{q} = 20 \, E_{\rm{R}}$ \cite{lattice_calibration} in each direction ($q=x,y,z$) with less than $4\%$ residual double occupancy. Here, $E_{\rm{R}}=\hbar^2 k_{\rm{l}}^2/(2 m)$ is the photon-recoil energy, with $k_{\rm{l}}=2 \pi / \lambda_{\rm{l}}$ and $m$ the mass of the Cs atom. The optical lattice results in a residual harmonic confinement of $\nu_z=11.9(2) \, \rm{Hz}$ in the $z$-direction of gravity. A broad Feshbach resonance allows us to set the atomic scattering length $a_{\rm{s}}$ and thus $U$ independently of $J$ by means of an offset magnetic field $B$ \cite{Mark2011}.

Tunneling resonances are observed by quickly tilting the lattice in the $z$-direction through a reduction of $|\nabla B|$ and then lowering $V_z$ to $10 \, E_{\rm{R}}$ within $1 \, \rm{ms}$, giving $J \approx 25$ Hz \cite{Jaksch1998,units}. All dynamics is now restricted along 1D Mott chains with an average length of 40 sites \cite{Meinert2013}. The chains, in total $\approx 2000$, are decoupled from each other on the relevant experimental timescales. We let the systems evolve for a hold time $t_{\rm{h}}$ of up to $200 \, \rm{ms}$ in the tilted configuration and then quickly ramp back $V_z$ to its original value and remove the tilt, thereby projecting the many-body state onto number states. The ensemble is characterized by measuring the number of doubly occupied sites $N_{\rm{d}}$ through Feshbach molecule formation with an overall efficiency of $80(3)\%$ \cite{Winkler2006,Danzl2010}. Alternatively, we detect the emergence of momentum-space coherence in time-of-flight (TOF) by quickly turning off all trapping potentials and allowing for $20 \, \rm{ms}$ of free levitated expansion at $a_{\rm{s}}=0$ \cite{Weber2002} before taking an absorption image.

The experimental result for a specific choice of $U=1077(20) \, \rm{Hz}$ is shown in Fig.~1 B. For a hold time of $t_{\rm{h}}=200 \, \rm{ms}$ the transient response as discussed below has settled to a steady-state value. Besides a broad resonance at $E = 1095(2) \, \rm{Hz}$ with FWHM $=172(9) \, \rm{Hz}$, two narrower resonances at $E = 532(1)$ and $351(1) \, \rm{Hz}$ with FWHM $=44(2)$ and $27(2) \, \rm{Hz}$ can be seen. While the broad resonance is the result of resonant tunnel coupling to nearest-neighbor dipole states at $E_1=U$ \cite{Meinert2013}, the positions of the narrower resonances are consistent with $E_2 = U/2$ and $E_3 = U/3$ and we hence interpret them to emerge from tunnel processes extending over a distance of two and three lattice sites, respectively. The reduced widths reflect the smaller amplitude of the higher-order tunnel processes. We believe that the resonances are slightly broadened inhomogeneously by the external harmonic confinement. The assignment of the resonance features to tunneling processes over multiple lattice sites is supported by TOF images (insets to Fig.~1 B) taken for each resonance $E_n$ in the course of the transient response. The images clearly exhibit matter-wave interference patterns, indicating delocalization of the atoms during the tunnel processes. The integrated line densities are presented in Figs.~1 (C)-(E). The periodicity of the sinusoidal density modulation, determined to $2 \hbar k_{\rm{l}}/n$, is in agreement with spatial coherence of the atomic wave function over a distance of $n$ sites.

We now investigate the transient dynamics following the quantum quench. Fig.~2 A and C (B and D) show the on-resonance response of $N_{\rm{d}}$ and the fringe visibility $V$ in the TOF images for $E_1$ ($E_2$). The quench to $E_1$ results in large amplitude oscillations for $N_{\rm{d}}$, which decay on a timescale of a few tens of ms. The dynamics for $E_2$ are highly overdamped and fit to a saturated growth function of the form $\propto (1-e^{-t_{\rm{h}}/\tau})$ with a characteristic rate $1/\tau$. Interestingly, on both resonances $N_{\rm{d}}$ relaxes to the same stationary value. The oscillations for $E_1$ at frequency $4J$ have been topic of our study in Ref.~\cite{Meinert2013}. The measured decay rates for the oscillatory response in $N_d$ are consistent with the width of the resonance in Fig.~1 B. Calculations show that the decay is due to many-body dephasing, which plays an increased role for larger chain lengths, and not due to trap imperfections, scattering losses, etc. \cite{Meinert2013}. The oscillatory response at $E_1$ is clearly reflected in the dynamics for $V$, as each local minimum coincides with an extremum for $N_{\rm{d}}$. For $E_2$ a simple three-site BH model predicts oscillations at frequency $\nu_2 = 4 (2\sqrt{2} + \sqrt{2}) J^2/U$ \cite{Methods}. In the experiment we find a single maximum for $V$ before it decays. The dephasing here results from more complicated dynamics in BH chains longer than three sites. In that case multiple atoms can tunnel, and inter-particle interactions give rise both to constraints on which atoms can tunnel simultaneously and to modifications of the tunneling amplitude \cite{Methods}. This results in processes with many competing frequency components, dephasing the oscillations in the doublon number. While the dephasing can be enhanced by inhomogeneities, e.g., due to the harmonic confinement, such contributions are small relative to the intrinsic dephasing in longer chains \cite{Methods}.

We now focus on the scaling of the resonant doublon growth rate $1/\tau$ with $J$ and $U$ for the resonance $E_2$. Example data sets, shown in Fig.~3 A, clearly demonstrate that $1/\tau$ depends not only on $V_z$ alone, but also on $U$ when $V_z$ and thereby $J$ is kept constant. In Fig.~3 B we plot the same data with the time axis rescaled by the energy scale $J^2/(U/2)$ for a second-order tunneling process. Remarkably, the data collapses onto a single curve, demonstrating that indeed second-order tunneling dominates the transient dynamics following the quench. The experimental data is supported by numerical simulations for 10-30 site BH chains \cite{Methods}. The numerical data shows the same rise characteristics and reveals the same scaling collapse, see Fig.~3 C. Next, we extract values for $1/\tau$ from measurements taken at different combinations of $V_z$ and $U$ and plot them in Fig.~3 D as a function of $J^2/(U/2)$. Our data reveals a linear dependence with a surprisingly large prefactor $\alpha_2 = 38(2)$, which we analyze in two ways. First, we compare to the frequency of coherent doublon oscillations in the simple three-site model. The role of many-body dephasing faster than a full second-order tunneling cycle is estimated by assuming $\tau$ as a quarter of the full tunneling period. The value $1/\tau \approx 4 \times \nu_2$ is indicated by the solid line in Fig.~3 D. Second, we extract a characteristic growth rate from the numerical data, indicated by the dashed regions in Figs.~3 (C) and (D), revealing very good quantitative agreement with the experiment.

Along the same lines we investigate the dynamical scaling of the resonant response at the resonance $E_3=U/3$. The dynamics that we observe for different combinations of $J$ and $U$ (Fig.~3 E) is qualitatively very similar to the one at $E_2$. The doublon number $N_{\rm{d}}$ settles to a steady-state value over a characteristic time $\tau$. In view of a third-order tunneling process we rescale the time axis by $J^3/(U/3)^2$ (Fig.~3 F). Again we find a striking collapse of the data. The result of our numerics is shown in Fig.~3 G. Also the numerical data collapses onto a single curve. Residual oscillations after the initial growth period relate to the finite system size and the lack of averaging over positions in the trap \cite{Methods}. In Fig.~3 H we plot $1/\tau$ as a function of $J^3/(U/3)^2$. Again the data collapses onto a single line. From the linear fit we obtain a slope of $\alpha_3=34(2)$. This is in good agreement with a characteristic growth rate determined from the numerical data, which we indicate by the dashed region as before. We note that the signature of the third-order process is not spoiled by the presence of second-order energy shifts \cite{Methods}.

Finally, in Fig.~4 A and B we show resonances corresponding to many-body tunneling across four and five lattice sites. For this data the lattice depth was reduced to $V_z = 8 \, E_{\rm{R}}$ and $7 \, E_{\rm{R}}$ to speed up the process while still assuring that the systems are initially in the Mott-insulating regime. With decreasing $V_z$ the resonances at $U/2$ and $U/3$ slightly broaden, which we attribute to the increase of the second- and third-order tunneling rate. The new resonances at $U/4$ and $U/5$ are weak but clearly detectable. We note that on these resonances the transient response is far from reaching a steady-state value, making it difficult to determine a characteristic time scale.

Our results underline the utility of cold atoms in optical lattices for the investigation of fundamental physical processes, even when these are driven by small-amplitude terms. By partly freezing the motion in the deep lattice, these sensitive processes can be observed here despite finite initial temperatures (which lead to defects and missing atoms). This and the long-range character of the higher-order tunnel coupling will further motivate investigation of quantum phases in spin models near these higher-order resonances, as well as further dynamical studies, including systems with tilts along multiple axes \cite{Sachdev2002,Pielawa2011}. It is also worth noting the parallels between the observed tunneling processes and multi-photon electron-positron creation in strong electric fields, which could open the door to quantum simulations of relativistic phenomena such as the Sauter-Schwinger effect in tilted Mott insulators \cite{Queisser2012}.


\bibliographystyle{Science}

\begin{scilastnote}
\item We are indebted to R. Grimm for generous support, and thank J. Schachenmayer for discussions and contributions to numerical code development. We gratefully acknowledge funding by the European Research Council (ERC) under Project No. 278417, and support in Pittsburgh from NSF Grant PHY-1148957.
\end{scilastnote}


\newpage
\clearpage

\begin{center}
\includegraphics[width=4cm]{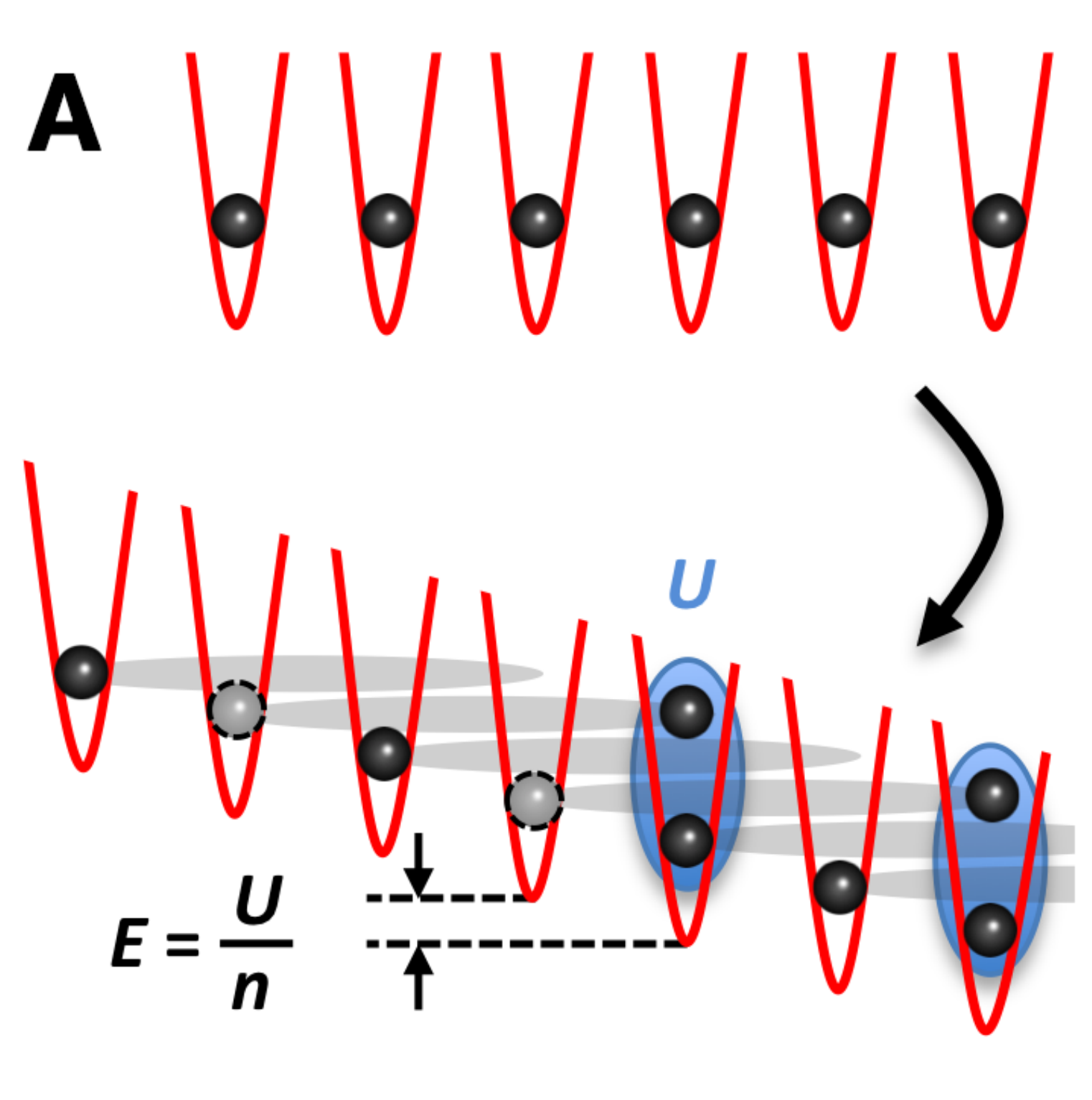}
\hspace{4mm}
\includegraphics[width=8cm]{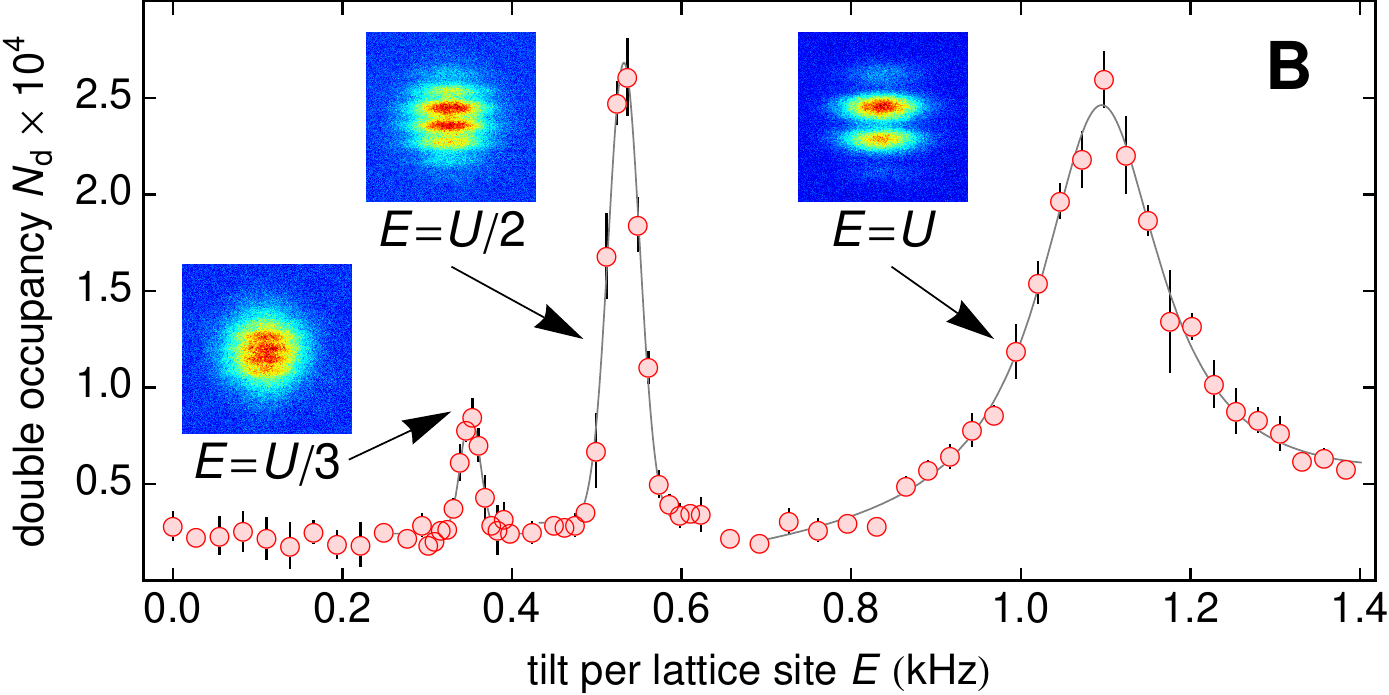}\\
\vspace{2mm}
\includegraphics[width=4cm]{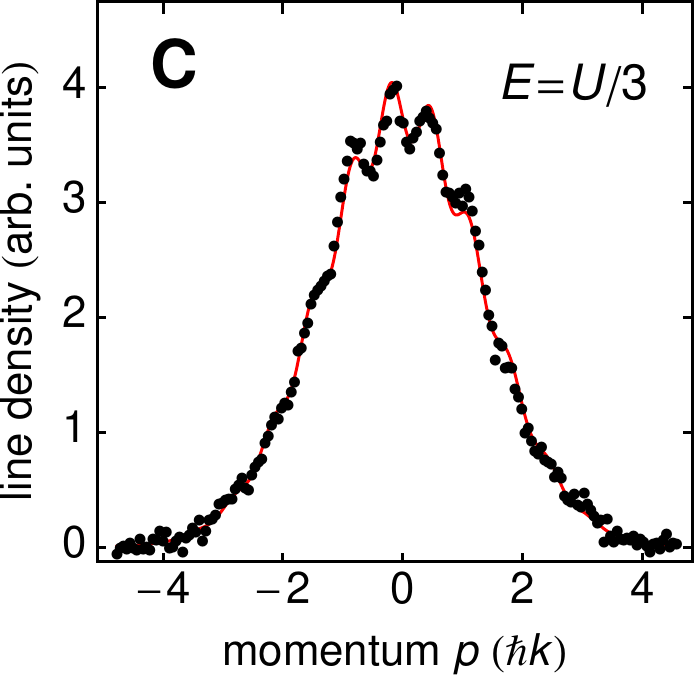}
\hspace{2mm}
\includegraphics[width=4cm]{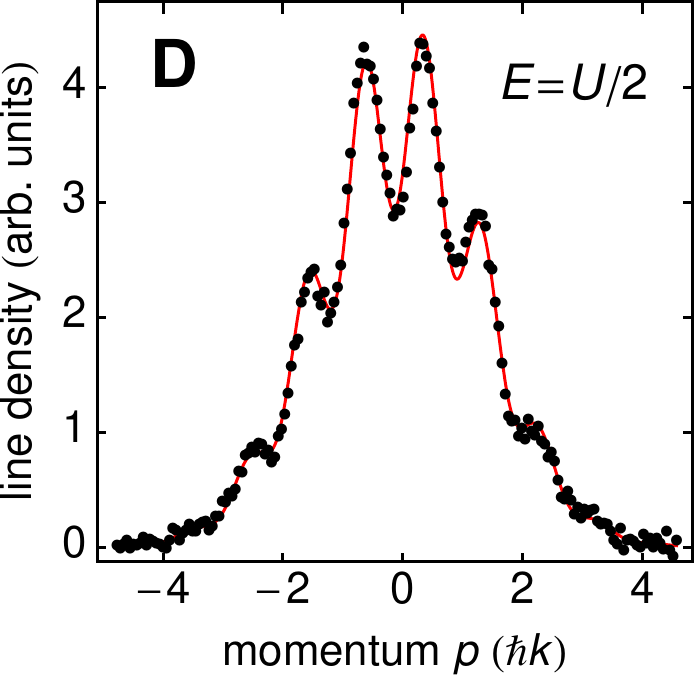}
\hspace{2mm}
\includegraphics[width=4cm]{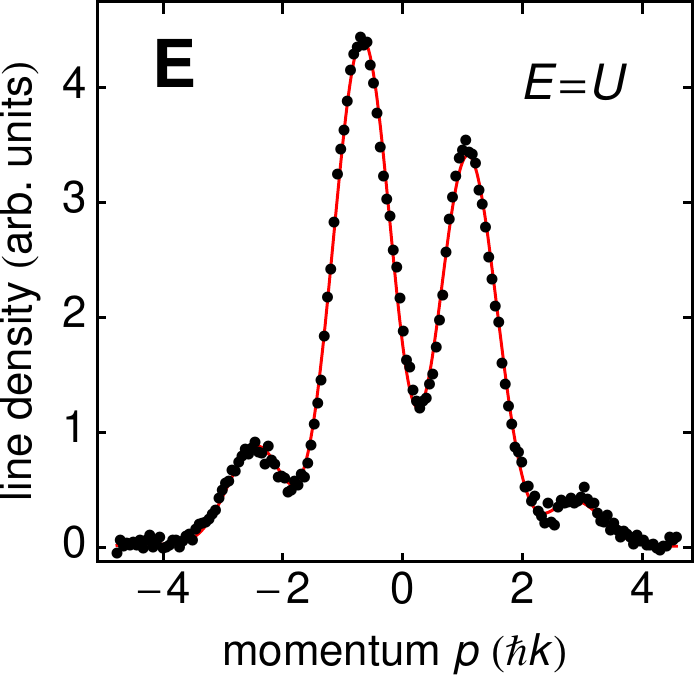}
\end{center}

\textbf{Fig. 1.} {\bf Tunneling resonances in a tilted 1D Mott insulator.} ({\bf A}) Schematic view of the long-range correlations across $n$ sites for a tilt of $E=U/n$ after the quench from the initial 1D one-atom Mott insulator (top) to the tilted configuration (bottom). Here, $n=3$ for illustration purposes. ({\bf B}) Number of doublons $N_{\rm{d}}$ as a function of $E$ for $t_{\rm{h}}=200 \, \rm{ms}$ after the quench. Here $V_z=10 \, E_{\rm{R}}$ and $a_{\rm{s}}=252(5) \, \rm{a_0}$, giving $U=1.077(20) \, \rm{kHz}$ for $V_{x,y}=20 \, E_{\rm{R}}$. The solid lines are Lorentzian (for $E=U$) and Gaussian (for $E=U/2$ and $E=U/3$) fits to the data to determine the center positions and widths. The insets show matter-wave interference patterns obtained in TOF at $E_1=U$, $E_2=U/2$, and $E_3=U/3$ taken after $t_{\rm{h}} = 1 \, \rm{ms}$, $9 \, \rm{ms}$, and $28 \, \rm{ms}$, respectively. ({\bf C})-({\bf E}) Integrated line densities of the TOF images shown in the insets in ({\bf B}). The solid lines are fits according to double slit interference patterns with Gaussian envelopes \cite{fitTOF}. Error bars in this and all other figures reflect the one-sigma standard deviation.

\newpage

\begin{center}
\includegraphics[width=5.76cm]{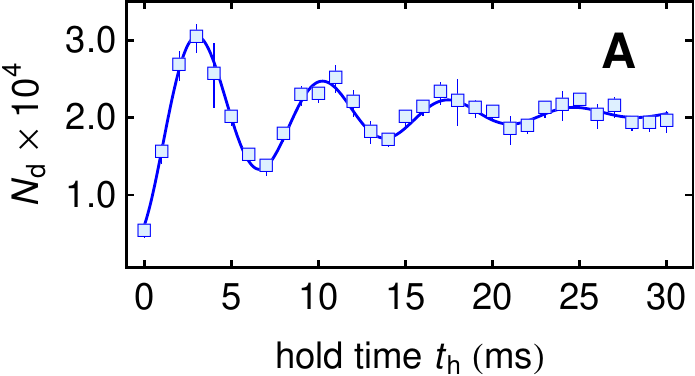}
\hspace{2mm}
\includegraphics[width=5.76cm]{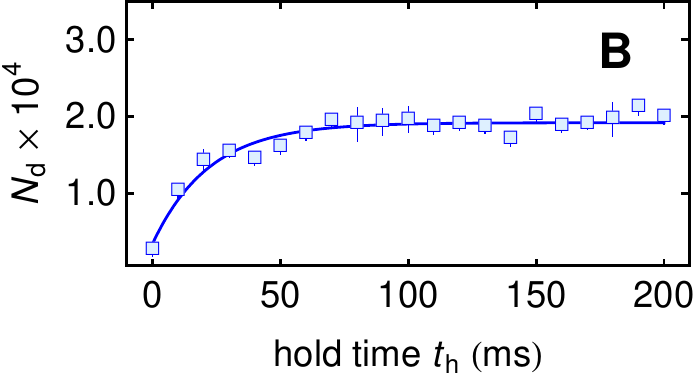}\\
\vspace{2mm}
\includegraphics[width=5.76cm]{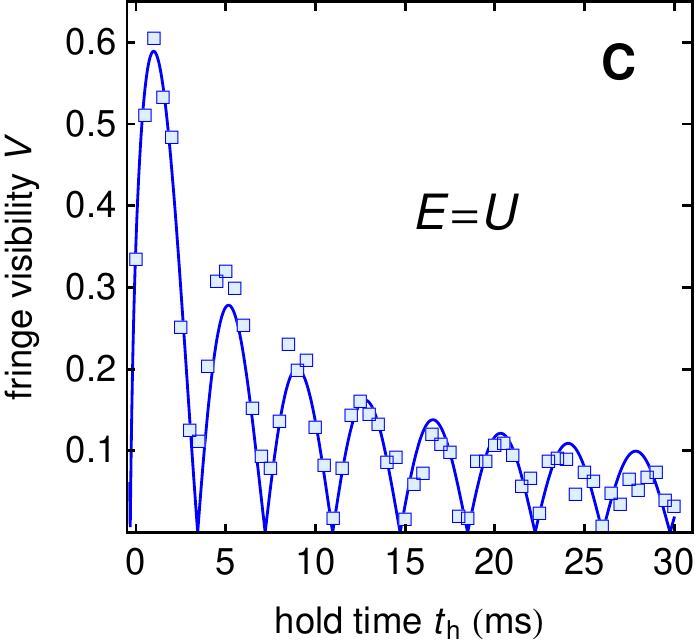}
\hspace{2mm}
\includegraphics[width=5.76cm]{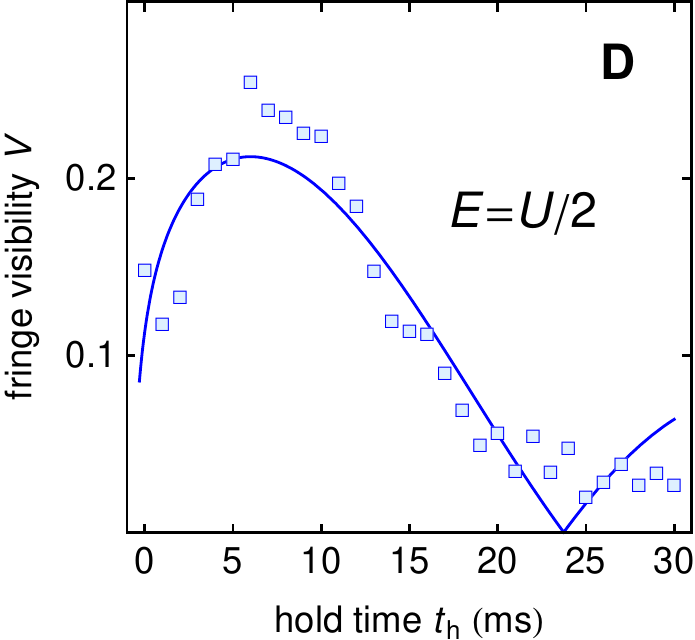}
\end{center}

\textbf{Fig. 2.} {\bf Comparison of the tunneling dynamics to nearest and second-nearest neighbors.} Double occupancy $N_{\rm{d}}$, ({\bf A}) and ({\bf B}), and fringe visibility $V$ in the TOF images, ({\bf C}) and ({\bf D}), as a function of hold time $t_{\rm{h}}$ after the quench. Coherent oscillations in $N_{\rm{d}}$ at $E_1=U$ in ({\bf A}) are contrasted to overdamped dynamics at $E_2=U/2$ ({\bf B}) for $V_z=10 \, E_{\rm{R}}$ and $a_{\rm{s}}=253(5) \, \rm{a_0}$. The evolution of $N_{\rm{d}}$ is fitted by an exponentially damped sinusoid and a saturated growth, respectively. The solid lines in ({\bf C}) and ({\bf D}) are fits to guide the eye based on the modulus of an algebraically decaying sinusoid.

\newpage

\newpage

\begin{center}
\includegraphics[width=3.9cm]{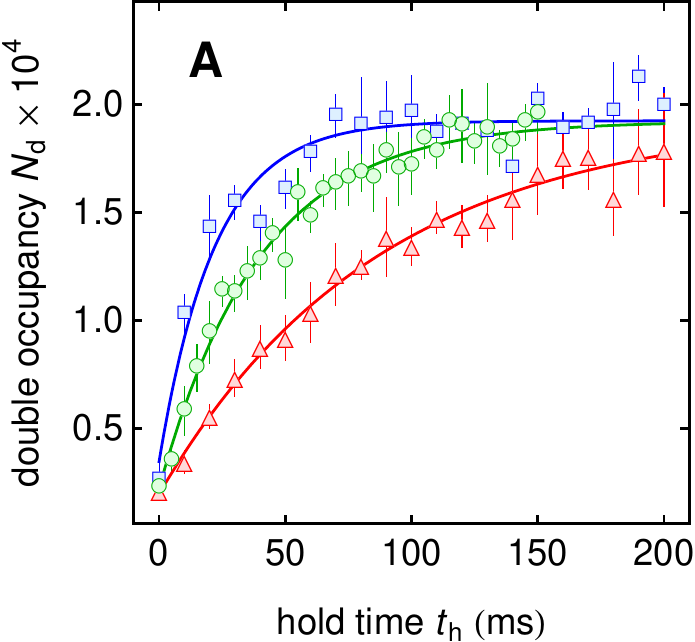}
\includegraphics[width=3.9cm]{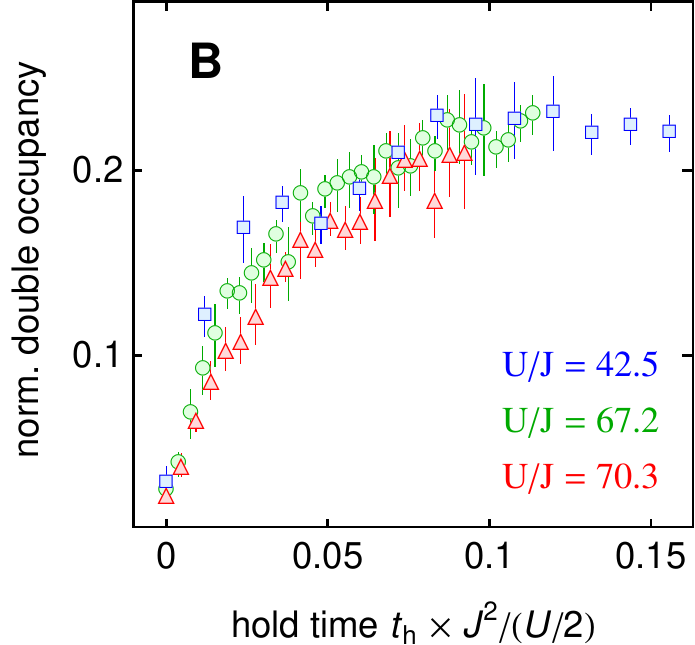}
\includegraphics[width=3.9cm]{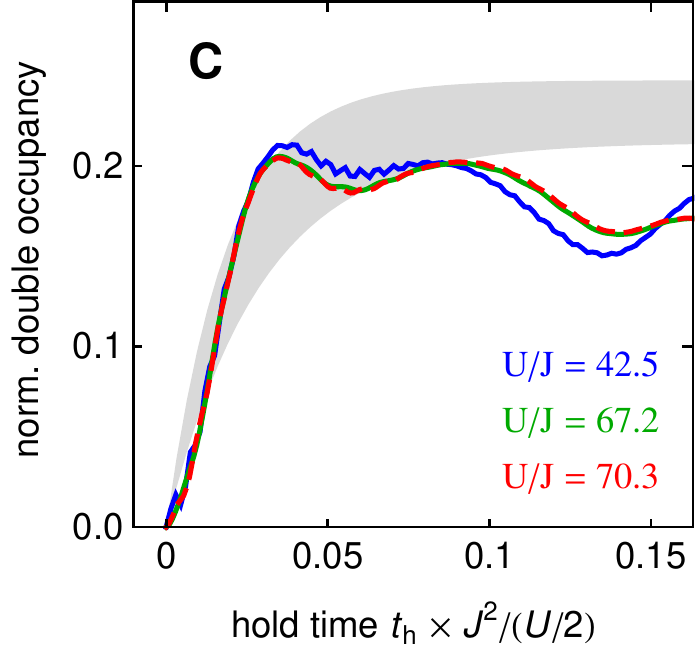}
\includegraphics[width=3.9cm]{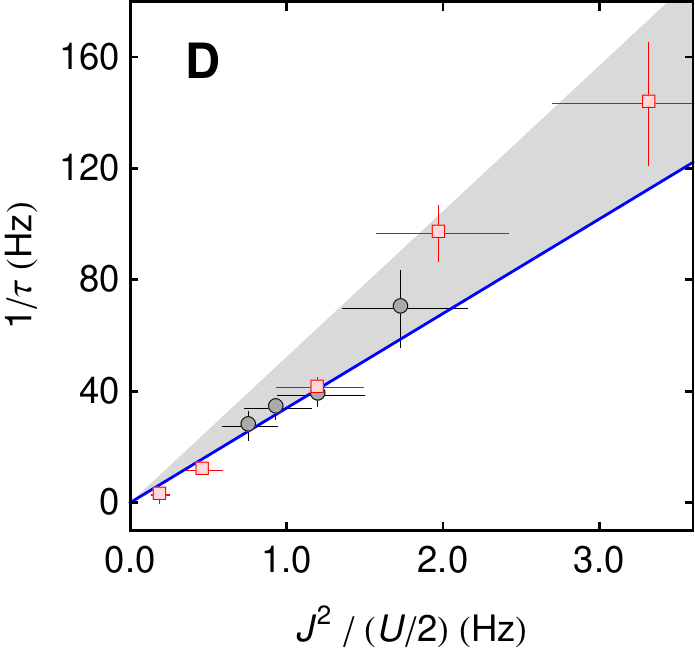}\\
\vspace{2mm}
\includegraphics[width=3.9cm]{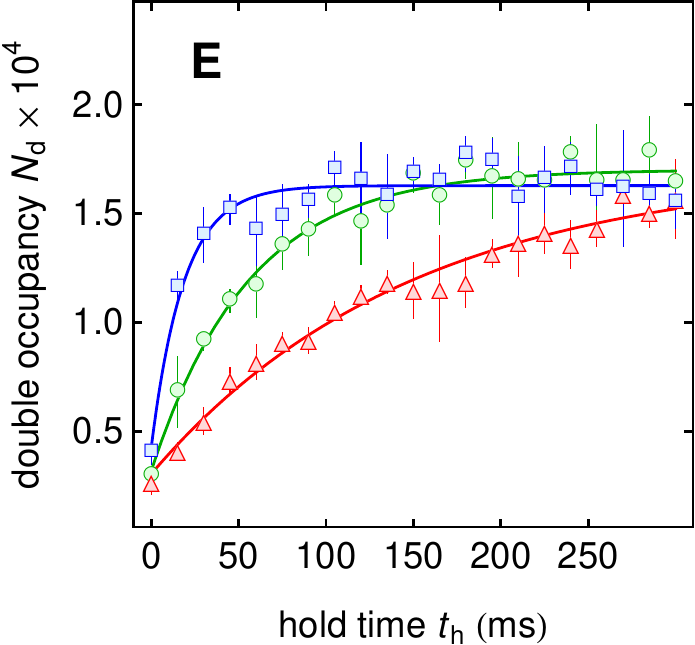}
\includegraphics[width=3.9cm]{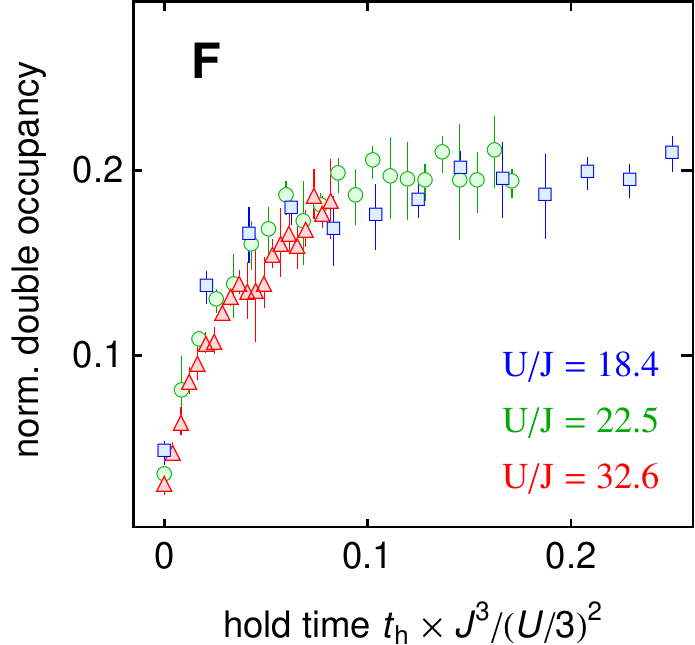}
\includegraphics[width=3.9cm]{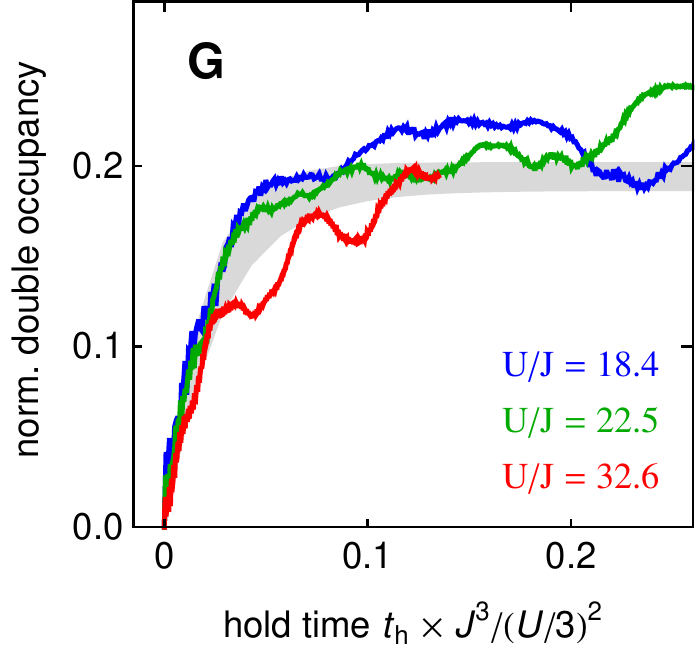}
\includegraphics[width=3.9cm]{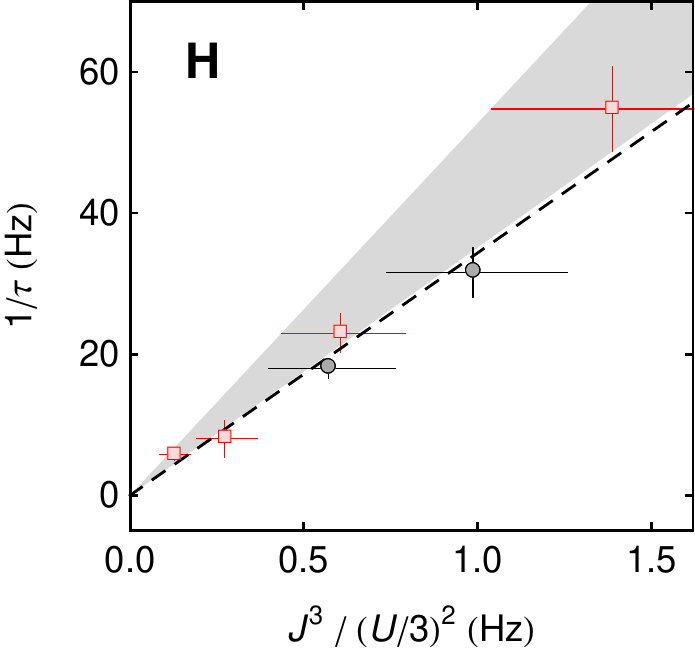}\\
\end{center}

\textbf{Fig. 3.} {\bf Second- and third-order tunneling dynamics.} Dependence of $N_{\rm{d}}$ on $J$ and $U$ at $E_2=U/2$, ({\bf A})-({\bf D}), and $E_3=U/3$, ({\bf E})-({\bf H}). ({\bf A}) Double occupancy at $E_2$ for $(V_z/E_{\rm{R}}, a_{\rm{s}}/\rm{a_0}) $ = $(10,253(5))$ (squares), $(12,253(5))$ (triangles), and  $(10,400(5))$ (circles). ({\bf E}) Double occupancy at $E_3$ for $(V_z/E_{\rm{R}}, a_{\rm{s}}/\rm{a_0}) $ = $(7,253(5))$ (squares), $(9,253(5))$ (triangles), and  $(9,175(5))$ (circles). The solid lines are fits to the data with saturated growth functions. ({\bf B}), ({\bf F}) Collapse of the data shown in ({\bf A}) and ({\bf E}) for rescaled time axes. ({\bf C}), ({\bf G}) Result of a numerical simulation of the resonant response at $E_2$ and $E_3$, respectively. ({\bf D}), ({\bf H}) Growth rates $1/\tau$ for $E_2$ and $E_3$, respectively. In ({\bf D}) the data for $V_z=(8,9,10,12,14) \, E_{\rm{R}}$ with fixed $a_{\rm{s}}=253 \, \rm{a_0}$ (squares) and $a_{\rm{s}} = (175,253,325,400) \, \rm{a_0}$ with fixed $V_z=10 \, E_{\rm{R}}$ (circles) is plotted as a function of $J^2/(U/2)$. The solid line gives the prediction from a three-site BH model (see text). In ({\bf H}) the data for $V_z=(7,8,9,10)  \, E_{\rm{R}}$ with $a_{\rm{s}}=253 \, \rm{a_0}$ (squares) and for $a_{\rm{s}}=175 \,  \rm{a_0}$ at $V_z=9 \, E_{\rm{R}}$ and $a_{\rm{s}}=300 \,  \rm{a_0}$ at $V_z=7 \, E_{\rm{R}}$ (circles) is plotted as a function of $J^3/(U/3)^2$. The dashed line is a linear fit to the experimental data. The shaded areas in ({\bf C}), ({\bf D}), ({\bf G}), and ({\bf H}) indicate the spread in the growth rate extracted from the numerical data with fixed steady-state values from the experiment \cite{Methods}.
\newpage

\begin{center}
\includegraphics[width=5.76cm]{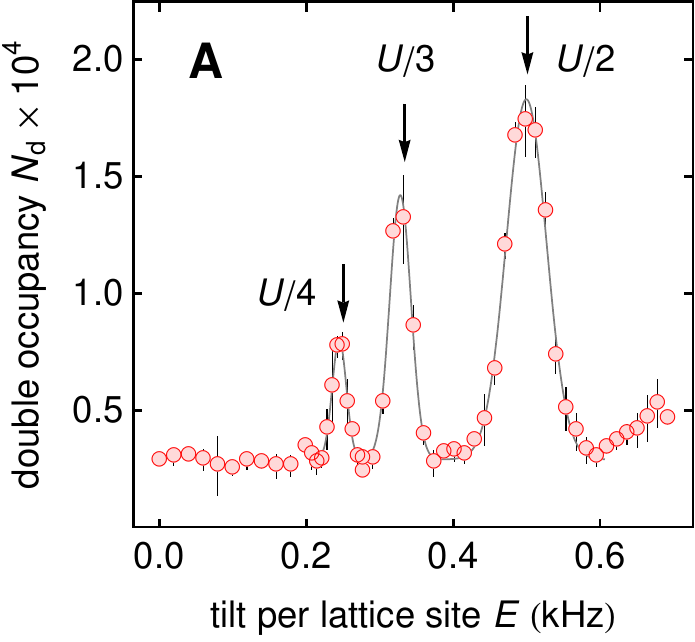}
\hspace{2mm}
\includegraphics[width=5.76cm]{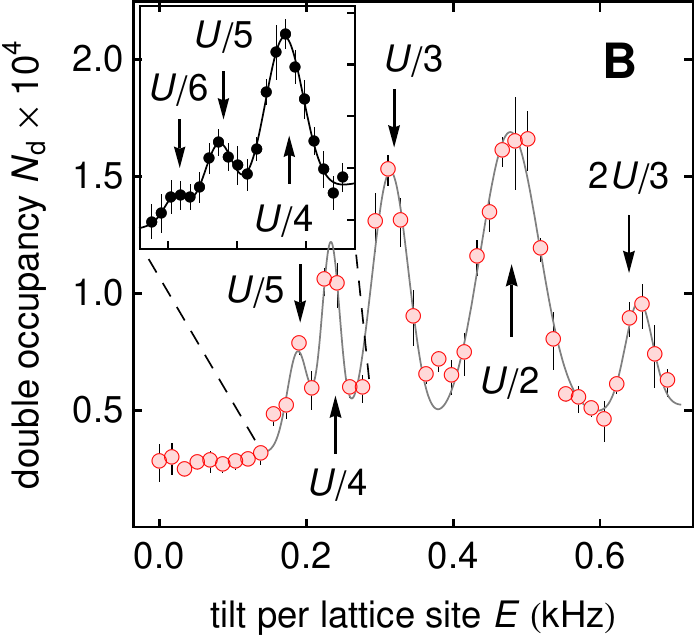}
\end{center}

\textbf{Fig. 4.} {\bf Higher-order tunneling resonances.} Double occupancy $N_{\rm{d}}$ as a function of $E$ after $t_{\rm{h}} = 200 \, \rm{ms}$ at ({\bf A}) $V_z=8 \, E_{\rm{R}}$ and ({\bf B}) $V_z=7 \, E_{\rm{R}}$ with $U = 1002(20) \, \rm{Hz}$ and $U = 959(20) \, \rm{Hz}$, respectively, for $a_{\rm{s}}=252(5) \, \rm{a_0}$. The arrows indicate the expected positions of the tunneling resonances at $E_n=U/n$. In ({\bf B}) an additional resonance at $2U/3$ appears. The inset to ({\bf B}) gives a fine scan of the $U/4$ and $U/5$ resonances. The solid lines in ({\bf A}) and ({\bf B}) are fits based on the sum of multiple Gaussians to guide the eye.

\newpage

\section{Supplementary Material for ``Observation of many-body long-range tunneling after a quantum quench''}

In this supplementary material, we provide more details of the theoretical treatment of the resonance behavior, including outlining the perturbation-theory models for the behavior near $E=U/n$, and also discussing the comparison between experimental results and data from numerical simulations.

\subsection{Perturbation theory solutions for the $E_2=U/2$ resonance}

In Ref.~\cite{Sachdev2002SuppMat}, a simplified spin model is derived describing the resonant dynamics near the $E=U$ resonance in our system. The dynamics seen here at the $E_n=U/n$ resonances can be considered in a similar fashion, by determining the corresponding resonant states and applying perturbation theory to determine the tunneling amplitudes and energy shifts.

Here we discuss this in more detail. As a simple introduction, we first discuss resonant oscillations in a three-site Hubbard model for the $E_2=U/2$ resonance, showing how the corresponding oscillation frequency is obtained. We then give an outline of the perturbation-theory description for multiple sites, and discuss the more complicated dynamics that arise in longer chains. In each case, we begin by identifying the manifold of atom configurations that have an equal energy (in the tilted system) to an initial state with a single atom on every lattice site, and that are connected by small numbers of processes in the lowest relevant order in perturbation theory. This is analogous to the starting point for the analysis of dynamics near the $E_1=U$ resonance in Ref.~\cite{Sachdev2002SuppMat}, and comprises the manifold of states that will be relevant in the dynamics at moderate times in a parameter regime where $J\ll U,E$.

\noindent \textbf{Simple case: Three lattice sites}

To the simplest approximation, we can discuss the dynamics at $E_2$ on the basis of a three-site Hubbard model. Starting with one atom per site, $|i \rangle = |1,1,1 \rangle$, we let the system evolve with $E=E_2=U/2$ for which it couples resonantly to the dipole state $|f\rangle = |2,1,0\rangle$. For $U \gg J$, the system is found to oscillate between $|i\rangle$ and $|f\rangle$ at a single frequency $\nu_2 = 2 (2\sqrt{2} + \sqrt{2}) J^2/(U/2)$. This result is most intuitively understood from second-order tunneling, which constitutes the dominant process due to energetically suppressed nearest-neighbor tunneling. The effective coupling between $|i\rangle$ and $|f\rangle$ arises in second-order perturbation theory via contributions from two virtual intermediate states
\begin{eqnarray}
\nu^{(1)} & = & \langle f | J \hat{a}_1^\dagger \hat{a}_2  | 1,2,0 \rangle \langle 1,2,0 | J \hat{a}_2^\dagger \hat{a}_3 | i \rangle / (U/2) \\
\nu^{(2)} & = & \langle f | J \hat{a}_2^\dagger \hat{a}_1  | 2,0,1 \rangle \langle 2,0,1 | J \hat{a}_1^\dagger \hat{a}_2 | i \rangle / (U/2) \, ,
\end{eqnarray}
giving $\nu_2 = 2 (\nu^{(1)} +\nu^{(2)})$. Here, $\hat{a}_i^\dagger$ ($\hat{a}_i$) are the bosonic creation (annihilation) operators at the $i$-th lattice site. The additional factor of 2 arises as $\nu_2$ gives the oscillation frequencies in the probabilities of finding the system in $|i\rangle$ or $|f\rangle$. As discussed below, there is also a small energy shift in second-order perturbation theory for three sites.\\

\begin{center}
\includegraphics[width=5.76cm]{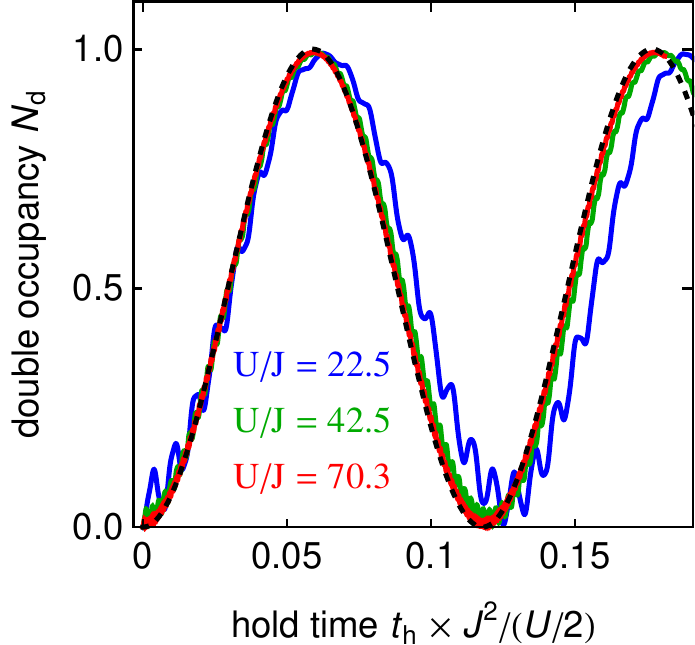}
\end{center}

\textbf{Fig. 1.} Double occupancy $N_{\rm{d}}$ as a function of hold time $t_{\rm{h}}$ for $E=U/2$ in a three-site Bose-Hubbard model at different values $U/J$. The dashed line shows the prediction in second-order perturbation theory for large $U/J$  giving a characteristic frequency $\nu_2$.\\

\noindent \textbf{General case: Occupation-dependent second-order tunneling }

For longer chains at the $E_2=U/2$ resonance, the configurations that are resonantly coupled involve situations where particles from the initial state have tunneled over two lattice sites, e.g., a configuration $ |1,1,1,1,1\rangle \rightarrow  |0,1,2,1,1\rangle$, or $ |1,1,1,1,1\rangle \rightarrow  |1,0,1,2,1\rangle$, or $ |1,0,1,2,1\rangle \rightarrow  |0,0,2,2,1\rangle$, and so forth. There is also a constraint that particles separated by two sites should not both tunnel simultaneously. In dynamics with chain lengths longer than 3 atoms, this constraint will produce a variety of frequency scales in the dynamics, and lead to a rapid dephasing of oscillations, as was observed for the $E_1=U$ resonance in Ref.~\cite{Meinert2013SuppMat}. The dephasing observed in the experiments at the $E_2$ and $E_3$ resonances is, however, even stronger than that observed at the $E_1$ resonance. This occurs because for the higher order resonances, interactions arise from additional sources.

The first additional source is that the sign and magnitude of the tunneling amplitude for any two-site tunneling process will depend on whether neighboring atoms (on either side) have tunneled. Let us consider the tunneling of the second atom in a chain, and look at the different amplitudes that can arise dependent on the position of the remaining atoms. Taking a general configuration of three atoms and considering the relevant amplitudes if we transfer a particle from the second site to the fourth site, we find the following possible amplitudes depending on the positions of the remaining atoms:
\begin{itemize}
\item $ |1,1,1,1,1\rangle\rightarrow |1,0,1,2,1\rangle$: There are two ways of obtaining this process in second order,  $ |1,1,1,1,1\rangle\rightarrow |1,0,2,1,1\rangle\rightarrow |1,0,1,2,1\rangle$
or $ |1,1,1,1,1\rangle\rightarrow |1,1,0,2,1\rangle\rightarrow |1,0,1,2,1\rangle$
so that we obtain an amplitude
$(2\sqrt{2}J^{2}+\sqrt{2}J^{2})/({-U/2})$, as discussed above.

\item $ |1,1,0,1,2\rangle\rightarrow |1,0,0,2,2\rangle$: There is one way of obtaining this term, with amplitude $(\sqrt{2}J^{2})/({U/2})$.

\item $ |0,1,2,1,1\rangle\rightarrow |0,0,2,2,1\rangle$: There are two ways of obtaining this,
$ |0,1,2,1,1\rangle\rightarrow |0,0,3,1,1\rangle\rightarrow |0,0,2,2,1\rangle$$ $
or $ |0,1,2,1,1\rangle\rightarrow |0,1,1,2,1\rangle\rightarrow |0,0,2,2,1\rangle$$ $, and we obtain an amplitude
$(3\sqrt 2 J^2)(-3U/2)+(2\sqrt 2 J^2)/(U/2)$

\end{itemize}
We can then also write a spin-model representation of the dynamics at this resonance, similar to that in Ref.~\cite{Sachdev2002SuppMat}, where spin up and down would now denote atoms having tunneled or not tunneled over two sites, respectively. Defining the standard Pauli matrices for a spin at position $l$ as $\sigma^x_l$, $\sigma^y_l$ and $\sigma^z_l$, we would not only obtain terms proportional to $\sigma^x_l$ terms that flip the spins, but in addition terms involving $\sigma^z_{l\pm 1} \sigma^x_{l}$ that give an interaction-dependent spin flip. \\

\noindent\textbf{Energy offset terms}

In addition to this, there are significant terms of order $J^{2}/U$ that arise in second order in perturbation theory and are diagonal in the occupation configuration basis.  For example, the energy of the state with occupation numbers $ |1,1,1,1,1\rangle$ is shifted compared with the unperturbed energy (at second order in the tunneling) by $-64J^{2}/(3U)$. When we investigate the energy values at second order in $J/U$ of states that have unperturbed energies (in the limit $J\rightarrow0$) equal to that of the state with occupations $ |1,1,1,1,1\rangle$, we find that the spread is comparable to the magnitude of the effective tunneling terms. Because of this, these terms lead to an exceptionally rapid dephasing of any oscillations in the dynamics of doubly occupied sites, which is seen in the calculations and measurements of the number of doubly occupied sites as a function of time.

\subsection{Higher-order resonances: $E_3=U/3$ and beyond}

Naturally, similar things will arise at $E_3= U/3$ and $E_4=U/4$, etc. In each case, the order of perturbation theory for the coupling terms will increase ($J^{3}/U^{2}$ for $E_3\approx U/3$ and $J^{4}/U^{3}$ for $E_4\approx U/4$), and the corresponding spin models will become more complicated. In each case, the diagonal shifts in second-order (i.e., $J^2/U$) terms will remain, and other diagonal corrections at third and higher order can enter, again leading to strong dephasing of oscillations and some suppression in the creation of doubly occupied sites, because the coupling terms are smaller than the energy shifts. We note that both in numerical simulations and in the experimental data, the rate of doublon growth for $E_3=U/3$ is still seen to be approximately proportional to $J^3/U^2$ despite the presence of the second-order energy shifts, because the tunneling that gives rise to growth in the doublon number is generated by the third-order terms.

Other resonances will also appear, e.g., for $E\approx2U/3$. We can see how this resonance arises at order $J^{3}/U^{2}$ if two particles tunnel, forming two doubly occupied sites, and one of them has tunneled just to the neighboring site, and another has tunneled over two sites. This given energy conservation, as the two doubly occupied sites cost us $2U$ in energy, but tunneling three sites down gains us $2U/3\times3=2U$.

\subsection{Numerical simulation of the behavior on resonance}

For the $E_2$ and $E_3$ resonances, we present numerical data for the number of doubly occupied sites as a function of time calculated within the Bose-Hubbard model augmented by a tilt \cite{Sachdev2002SuppMat}
\begin{equation}
\hat{H} = -J \sum\limits_{\langle i,j \rangle } \hat{a}_i^\dagger \hat{a}_j + \sum\limits_{i} \frac{U}{2} \hat{n}_i\left(\hat{n}_{i}-1\right) +E \sum\limits_i i \hat{n}_{i}  + \epsilon \sum\limits_i i^2 \hat{n}_{i} \, ,
\label{EQ1}
\end{equation}
where $\hat{n}_i = \hat{a}_i^\dagger \hat{a}_i$ are the number operators and $\epsilon$ accounts for a weak harmonic confinement. These are computed using exact diagonalization (ED) for small system sizes and long times, or using time-dependent density matrix renormalization group (t-DMRG) methods \cite{Vidal04SuppMat,tdmrg1SuppMat,tdmrg2SuppMat,dmrgrevSuppMat} for longer system sizes but shorter times. In each case, the computations presented were converged in the numerical parameters including time steps and t-DMRG truncation error. We calculated the propagation of the system in time, beginning with an initial state with one particle on each lattice site.

For the initial growth region, we found that beyond a very small number of particles and lattice sites (ca. 6 sites), the initial growth of doublons for $E_2=U/2$ and $E_3=U/3$ collapses essentially on a single line when scaled with $J^2/(U/2)$ for the $E_2$ resonance and with $J^3/(U/3)^2$ for the $E_3$ resonance. In the main article this was only shown for small system sizes. In Fig.~2 we show the same results for short times, now including t-DMRG simulations for 30 particles on 30 lattice sites, showing that this does not change in longer chains.\\

\begin{center}
\includegraphics[width=5.76cm]{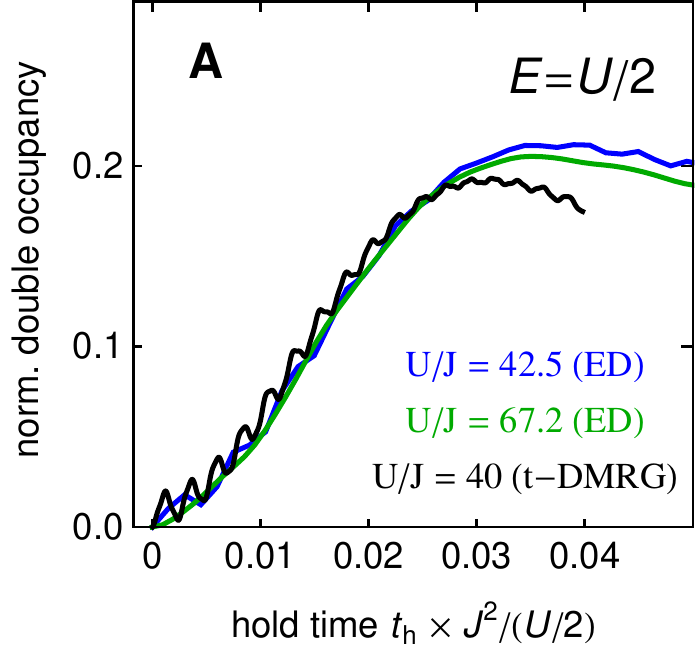}
\hspace{2mm}
\includegraphics[width=5.76cm]{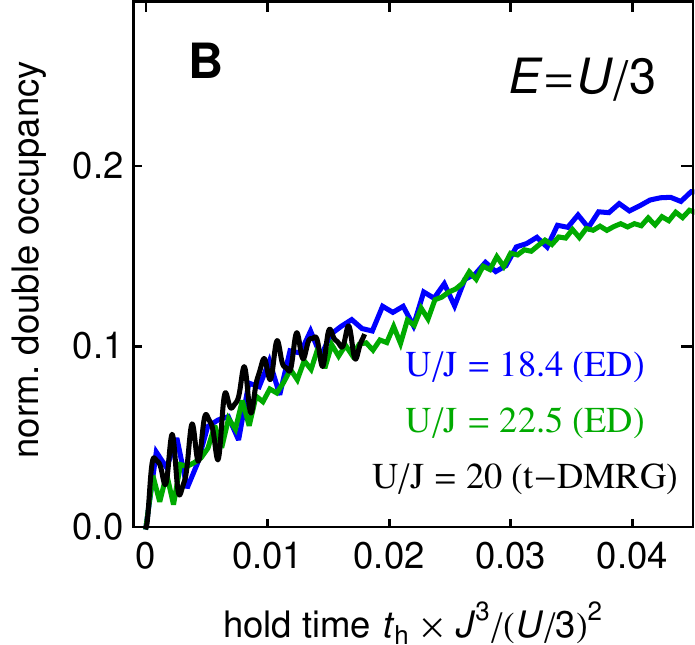}
\end{center}

\textbf{Fig. 2.} Initial growth in the number of doublons normalized to the total number of particles for $E=U/2$ ({\bf A}) and $E=U/3$ ({\bf B}) when increasing the system size. Numerical data for a 10 site Bose-Hubbard chain computed via exact diagonalization (blue and green) is compared to a system with 30 particles on 30 sites obtained by t-DMRG methods (black). The simulations include a weak harmonic confinement $\epsilon / J = 0.01$ for $E=U/2$ and $\epsilon / J = 0.005$ for $E=U/3$.\\

The key noticeable discrepancy between the data presented in Fig.~3 of the main article and the numerical simulations is the substantial oscillations in the numerical data after the initial growth region. While the dynamics of these oscillations is not the main subject of this article, let us comment briefly on these based on our numerical calculations and the relationship to the experimental data. These oscillations are clearly much more pronounced for smaller system sizes, and we find that they become smaller (though not non-existent) for longer chains in our calculations. In the experiment, where many chains are measured in parallel there are three types of averages taking place automatically that are not accounted for in the simulations, each of which lead to strong suppression of the oscillations under averaging in our calculations.

The first is an average over chain length, because the trapping potential confining the system leads to different chain lengths in different parts of the system. The second is averages over position within the harmonic trap, and the third is an average over positions of missing atoms in the initial state. We have simulated each of these contributions for small system sizes around 10 lattice sites, and find that the exact form of the oscillations - including their position and amplitude - depends heavily on the choice of trap strength, position within the harmonic trap, and position of missing atoms in the chain. To give an example of the effects of averaging, in Fig.~3, we plot the same results as in Fig.~3 of the main article, but now averaging the numerical data over trap positions. Here we simply shift the trap either one or two sites up or one or two sites down the gradient potential, and average the results from the resulting five trap positions. It is clear from the figure that for the $U/3$ resonance, this almost completely suppresses the oscillations.\\

\begin{center}
\includegraphics[width=5.76cm]{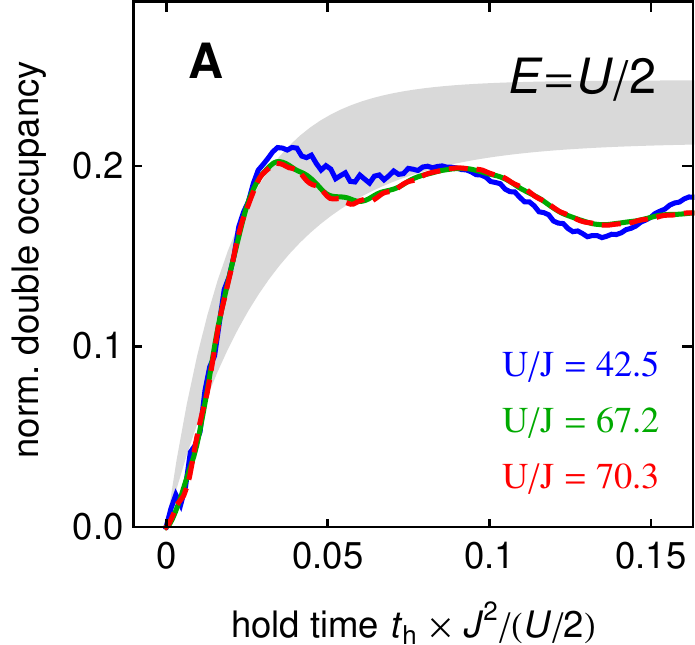}
\hspace{2mm}
\includegraphics[width=5.76cm]{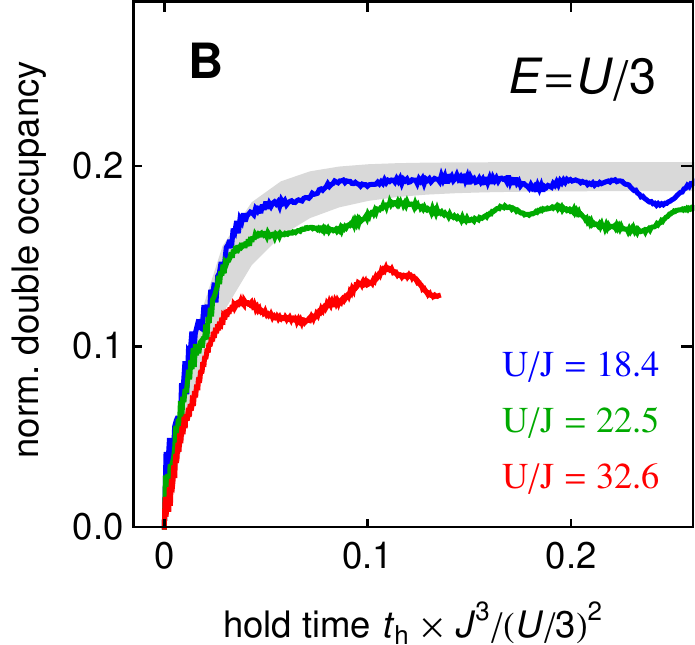}
\end{center}

\textbf{Fig. 3.} Number of doublons normalized to the total number of particles as a function of hold time $t_{\rm{h}}$ for $E=U/2$ ({\bf A}) and $E=U/3$ ({\bf B}) when averaging over trap positions. The harmonic confinement is set to $\epsilon / J = 0.01$ for both $E=U/2$ and $E=U/3$. The grey shaded areas indicate results of the fit method to the numerical data used to extract estimates of the characteristic growth rate $1/\tau$.\\

\noindent\textbf{Extracting estimates for the growth rate $1/\tau$ from the numerical data}

In the main article, we extract growth rates $1/\tau$ from the numerical simulations shown in Figs.~3 (C) and (G) and compare them to our experimental data in Figs.~3 (D) and (H). Specifically, we fit the initial increase of doubly occupied sites in the numerical data with a saturated growth function $N_{\rm{d}}^{\rm{max}} (1-e^{-t_{\rm{h}}/\tau})$ while fixing the steady-state value $N_{\rm{d}}^{\rm{max}}$ to what we obtain in our experimental data. The uncertainty in the obtained values for $1/\tau$ mainly reflects experimental fluctuations in $N_{\rm{d}}^{\rm{max}}$ and is indicated by the grey shaded regions.


\begin{thebibliography}{scibib}
\bibitem{Methods} Materials, methods, and additional theoretical background are available as supporting material on Science Online.
\bibitem{lattice_calibration} The lattice depth $V_q$ is calibrated via Kapitza-Dirac diffraction. The statistical error for $V_q$ is 1\%, though the systematic error can reach up to 5\%.
\bibitem{units} We give all energies in frequency units.
\bibitem{fitTOF} The matter-wave interference pattern is fit by $n(p) = N_0 e^{-\left( p-p_0 \right)^2/w^2} \left( 1 + V cos[k(p-p_0)+\phi]\right)$ with the fringe visibility $V$, the wave vector $k$, and a phase $\phi$.
\end{thebibliography}
\end{document}